\documentclass[twocolumn]{aastex63}
\usepackage{appendix}
\usepackage{longtable}
\usepackage{amsmath}
\usepackage{chemformula}

\shorttitle{First detection of \ch{NH2C(O)CH2OH} in the ISM}
\shortauthors{Rivilla et al.}

\begin{document}

\title{First glycine isomer detected in the interstellar medium: glycolamide (\ch{NH2C(O)CH2OH})}

\author[0000-0002-2887-5859]{V\'ictor M. Rivilla}
\affiliation{Centro de Astrobiolog{\'i}a (CAB), INTA-CSIC, Carretera de Ajalvir km 4, Torrej{\'o}n de Ardoz, 28850 Madrid, Spain}
\author[0000-0001-9629-0257]{Miguel Sanz-Novo}
\affiliation{Centro de Astrobiolog{\'i}a (CAB), INTA-CSIC, Carretera de Ajalvir km 4, Torrej{\'o}n de Ardoz, 28850 Madrid, Spain}
\affiliation{Grupo de Espectroscopia Molecular (GEM), Edificio Quifima, Laboratorios de Espectroscopia y Bioespectroscopia, Parque Cient\'ifico UVa, Universidad de Valladolid, 47011 Valladolid, Spain}
\affiliation{Computational Chemistry Group, Departamento de Química Física y Química Inorgánica, Universidad de Valladolid, E-47011 Valladolid, Spain}
\author[0000-0003-4493-8714]{Izaskun Jim\'enez-Serra}
\affiliation{Centro de Astrobiolog{\'i}a (CAB), INTA-CSIC, Carretera de Ajalvir km 4, Torrej{\'o}n de Ardoz, 28850 Madrid, Spain}
\author[0000-0003-4561-3508]{Jes\'us Mart\'in-Pintado}
\affiliation{Centro de Astrobiolog{\'i}a (CAB), INTA-CSIC, Carretera de Ajalvir km 4, Torrej{\'o}n de Ardoz, 28850 Madrid, Spain}
\author[0000-0001-8064-6394]{Laura Colzi}
\affiliation{Centro de Astrobiolog{\'i}a (CAB), INTA-CSIC, Carretera de Ajalvir km 4, Torrej{\'o}n de Ardoz, 28850 Madrid, Spain}
\author[0000-0003-3721-374X]{Shaoshan Zeng}
\affiliation{Star and Planet Formation Laboratory, Cluster for Pioneering Research, RIKEN, 2-1 Hirosawa, Wako, Saitama, 351-0198, Japan}
\author[0000-0002-6389-7172]{Andr\'es Meg\'ias}
\affiliation{Centro de Astrobiolog{\'i}a (CAB), INTA-CSIC, Carretera de Ajalvir km 4, Torrej{\'o}n de Ardoz, 28850 Madrid, Spain}
\author[0000-0001-6049-9366]{\'Alvaro L\'opez-Gallifa}
\affiliation{Centro de Astrobiolog{\'i}a (CAB), INTA-CSIC, Carretera de Ajalvir km 4, Torrej{\'o}n de Ardoz, 28850 Madrid, Spain}
\author[0000-0001-5191-2075]{Antonio Mart\'inez-Henares}
\affiliation{Centro de Astrobiolog{\'i}a (CAB), INTA-CSIC, Carretera de Ajalvir km 4, Torrej{\'o}n de Ardoz, 28850 Madrid, Spain}
\author[0000-0002-7387-9787]{Sarah Massalkhi}
\affiliation{Centro de Astrobiolog{\'i}a (CAB), INTA-CSIC, Carretera de Ajalvir km 4, Torrej{\'o}n de Ardoz, 28850 Madrid, Spain}
\author[0000-0002-4782-5259]{Bel\'en Tercero}
\affiliation{Observatorio Astron\'omico Nacional (OAN-IGN), Calle Alfonso XII, 3, 28014 Madrid, Spain}
\author[0000-0002-5902-5005]{Pablo de Vicente}
\affiliation{Observatorio de Yebes (OY-IGN), Cerro de la Palera SN, Yebes, Guadalajara, Spain}
\author[0000-0001-9281-2919]{Sergio Mart\'in}
\affiliation{European Southern Observatory, Alonso de C\'ordova 3107, Vitacura 763 0355, Santiago, Chile}
\affiliation{Joint ALMA Observatory, Alonso de C\'ordova 3107, Vitacura 763 0355, Santiago, Chile}
\author[0000-0001-7535-4397]{David San Andr\'es}
\affiliation{Centro de Astrobiolog{\'i}a (CAB), INTA-CSIC, Carretera de Ajalvir km 4, Torrej{\'o}n de Ardoz, 28850 Madrid, Spain}
\author[0009-0009-5346-7329]{Miguel A. Requena-Torres}
\affiliation{University of Maryland, College Park, ND 20742-2421 (USA)}
\affiliation{Department of Physics, Astronomy and Geosciences, Towson University, Towson, MD 21252, USA}
\author[0000-0002-3146-8250]{Jos\'e Luis Alonso}
\affiliation{Grupo de Espectroscopia Molecular (GEM), Edificio Quifima, Laboratorios de Espectroscopia y Bioespectroscopia, Parque Cient\'ifico UVa, Universidad de Valladolid, 47011 Valladolid, Spain}




\begin{abstract}
We report the first detection in the interstellar medium of a \ch{C2H5O2N} isomer: $syn$-glycolamide (\ch{NH2C(O)CH2OH}). The exquisite sensitivity at sub-mK levels of an ultra-deep spectral survey carried out with the Yebes 40m and IRAM 30m telescopes towards the G+0.693-0.027 molecular cloud have allowed us to unambiguously identify multiple transitions of this species.
We derived a column density of (7.4 $\pm$ 0.7)$\times$10$^{12}$\,cm$^{-2}$, which implies a molecular abundance with respect to \ch{H2} of 5.5$\times$10$^{-11}$. The other \ch{C2H5O2N} isomers, including the higher-energy $anti$ conformer of glycolamide, and two conformers of glycine, were not detected. The upper limit derived for the abundance of glycine indicates that this amino acid is surely less abundant than its isomer glycolamide in the ISM. The abundances of the \ch{C2H5O2N} isomers cannot be explained in terms of thermodynamic equilibrium, and thus chemical kinetics need to be invoked. While the low abundance of glycine might not be surprising, based on the relative low abundances of acids in the ISM compared to other compounds (e.g. alcohols, aldehydes or amines), several chemical pathways can favour the formation of its isomer glycolamide. 
It can be formed through radical-radical reactions on the surface of dust grains. The abundances of these radicals can be significantly boosted in an environment affected by a strong ultraviolet field induced by cosmic rays, such as that expected in G+0.693-0.027. Therefore, as shown by several recent molecular detections towards this molecular cloud, it stands out as the best target to discover new species with carbon, oxygen and nitrogen with increasing chemical complexity.

\end{abstract}







\section{Introduction}
\label{sec:intro}

Glycine (or 2-aminoacetic acid, \ch{NH2CH2COOH}) is the simplest representative of amino acids, the building blocks of proteins, which play fundamental catalytic and metabolic roles in biochemistry. Given that some fundamental molecular precursors that contributed to the origin of life on Earth could have an extraterrestrial origin (e.g.~\citealt{pearce2017}), numerous efforts have been focused on the search for glycine in space in the last decades. Along with other amino acids, glycine has been detected in chondritic meteorites (\citealt{cronin1983,burton2012}), and very recently in samples provided by the JAXA Hayabusa2 mission of the asteroid Ryugu (\citealt{Potiszil2023}). 
It was also detected in the comet 67P/Churyumov-Gerasimenko  by the {\it in-situ} mass spectrometric analysis performed by the ESA spacecraft Rosetta (\citealt{altwegg2016}). These detections put forward the possibility of glycine to be originally formed in the interstellar medium (hereafter ISM) and then incorporated to the protosolar nebula (\citealt{pizzarello2006}). 
In the context of this latter hypothesis, glycine, along more complex amino acids and other prebiotic chemicals, might have been delivered to early Earth through cometary and meteoritic impacts (\citealt{oro1961_comet,cronin1988,chyba1990,chyba1992,cooper2001,pasek2008,glavin2009,pearce2017,marty2017,obrien2018,rubin2019,rivilla2020a}).

Laboratory experiments have shown that glycine can be synthesized under interstellar conditions (e.g.~\citealt{munoz-caro2002,ioppolo2021}). \citet{kuan2003} claimed the presence of glycine towards several high-mass star-forming regions, although a rigorous verification performed by \citet{snyder2005} finally disproved these detections. Glycine has not been detected yet in the ISM, despite the multiple observational efforts targeting different astronomical environments (e.g. \citealt{combes1996,ceccarelli_structure_2000,snyder2005,jones2007,cunningham2007,jimenez-serra_spatial_2016,jimenez-serra2020}). 

The non detection of glycine might imply that it is not the most abundant member of the \ch{C2H5O2N} isomeric family in the ISM. 
Indeed, glycine is not the most thermodynamically stable isomer, but two other isomers (\citealt{sanz-novo2019}): N-methylcarbamic acid (\ch{CH3NHCOOH}), whose microwave spectroscopy is not available, and $syn-$methyl carbamate (\ch{CH3OC(O)NH2}), whose search has been unfruitful so far (\citealt{demyk2004}). Other isomers less stable than glycine, such as glycolamide (\ch{NH2C(O)CH2OH}; \citealt{sanz-novo2020,colzi2021}), and N-(Z)-hydroxyacetamide (\ch{CH3C(O)NHOH}; \citealt{sanz-novo2022b}) remain elusive.

In this work, we have searched for all the \ch{C2H5O2N} isomers for which microwave rotational spectroscopy is currently available, towards the chemically rich G+0.693-0.027 molecular cloud (G+0.693 hereafter), located in the Sgr B2 region of the Galactic Center. Many prebiotically relevant molecules have been detected for the first time in the ISM towards this source in the last four years, which makes it a very promising candidate for new discoveries (\citealt{rivilla2019b,rivilla2020b,rivilla2021a,rivilla2021b,rivilla2022a,rivilla2022b,rodriguez-almeida2021a,rodriguez-almeida2021b,zeng2021,jimenez-serra2022}).  
As a result of our search, we report the first detection in the ISM of a member of the \ch{C2H5O2N} isomeric family: the $syn$ conformer of glycolamide (\ch{NH2C(O)CH2OH}), as well as very constraining upper limits of the abundances for the other isomers, including two conformers of glycine. 

\begin{figure*}
\begin{center}
\includegraphics[width=18cm]{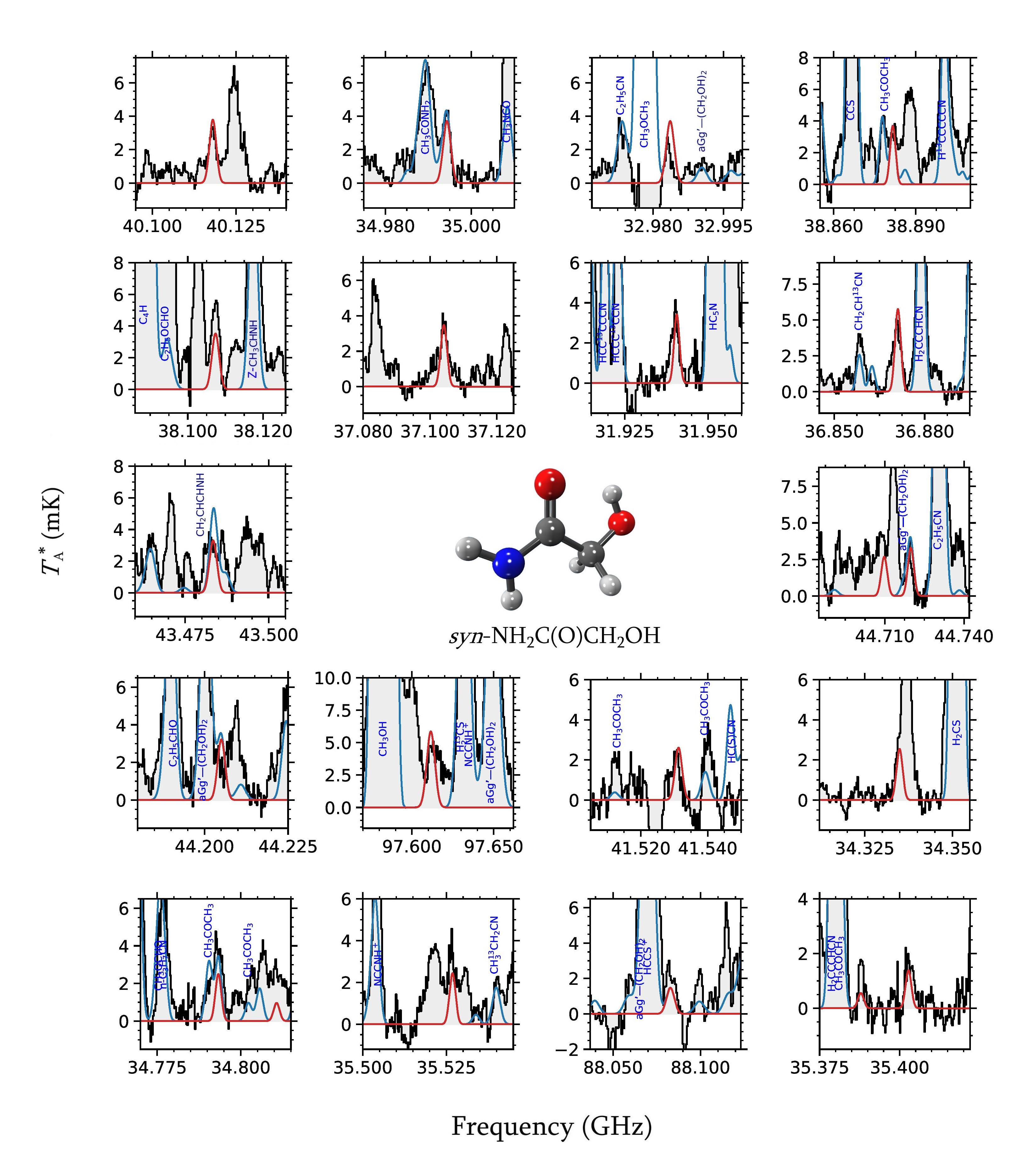}
\vspace{-15mm}
\end{center}
\caption{Unblended or slightly blended transitions of glycolamide detected towards the G+0.693, displayed in order of decreasing line intensity. The observed spectra are shown as a grey histogram. The best LTE fit derived with \textsc{Madcuba} for the $syn-$glycolamide emission is shown with a red curve. The blue curve represents the total modelled emission considering all the species identified towards this molecular cloud. In the center, the molecular structure of the $syn$ conformer of glycolamide is shown. Carbon atoms in grey, oxygen atoms in red, nitrogen atom in blue, and hydrogen atoms in white.}
\label{fig-spectra}
\end{figure*}

\section{Observations} 
\label{sec:observations}
 
We have significantly improved the sensitivity of the spectral survey towards G+0.693 used in previous works (e.g. \citealt{zeng2020,rivilla2021a,rivilla2021b,rivilla2022a,rivilla2022b,rivilla2022c,rodriguez-almeida2021a,rodriguez-almeida2021b,zeng2021,colzi2022,jimenez-serra2022,zeng2023}), performing new ultra-deep observations with the Yebes 40m (Guadalajara, Spain) and IRAM 30m (Granada, Spain) telescopes. 
The observations, using position switching mode, were centered at $\alpha$(ICRS) =$\,$17$^{\rm h}$47$^{\rm m}$22$^{\rm s}$, $\delta$(ICRS) =$\,-$28$^{\circ}$21$^{\prime}$27$^{\prime\prime}$, with the off position shifted by $\Delta\alpha$~=~$-885$$^{\prime\prime}$ and $\Delta\delta$~=~$290$$^{\prime\prime}$.
The line intensity of the spectra was measured in units of antenna temperature ($T_{\mathrm{A}}^{\ast}$) since the molecular emission towards G+0.693 is extended beyond the telescope primary beams (\citealt{requena-torres_organic_2006,requena-torres_largest_2008,zeng2018,zeng2020}).

The Yebes 40m observations (project 21A014; PI Rivilla) were carried out
during multiple sessions between March 2021 and March 2022, with a total telescope scheduled time of 230\,hours, of which 110\,hours were on source.
Part of this project, based on half of the observing runs, was presented in \citet{rivilla2022a,rivilla2022b}.
The Nanocosmos Q-band (7\,mm) HEMT receiver was used, which provides ultra broad-band observations (18.5\,GHz) in two linear polarizations (\citealt{tercero2021}). The receiver was connected to 16 FFTS providing a channel width of 38\,kHz. 
We observed two different spectral setups centered at 41.4 and 42.3\,GHz, with a total frequency coverage of 31.07$-$50.42\,GHz. 
Telescope pointing and focus were checked every one or two hours through pseudo-continuum observations.
The half power beam width (HPBW) of the telescope was $\sim$35$^{\prime\prime}$$-$55$^{\prime\prime}$ in the observed frequency range.
We performed an initial data inspection and reduction using a Python-based script\footnote{\url{https://github.com/andresmegias/gildas-class-python/}} that uses the \textsc{Class} module of the \textsc{Gildas} package (\citealt{megias2023}).
For each observing day, we automatically fitted baselines using an iterative method that first masks the more visible lines using sigma-clips, and then applies rolling medians and averages, interpolating the masked regions with splines. Spectra were then combined, averaged, and imported to \textsc{Madcuba} (\citealt{martin2019}).
The comparison of the spectra of the two frequency setups was used to identify possible spurious signals and other technical artifacts. 
The final spectra were smoothed to 256\,kHz, which translates into velocity resolutions of 1.5$-$2.5 km s$^{-1}$ in the range observed, which are enough to properly characterize the typical linewidths of 15$-$20 km s$^{-1}$ observed in the region (e.g. \citealt{requena-torres_organic_2006,zeng2018,rivilla2022c}).
We have checked flux density consistency between the new observations and the previous Yebes survey (e.g. \citealt{zeng2020,rodriguez-almeida2021a}). We find that the line intensities are consistent within a 5$\%$ level. The achieved noise at this spectral resolution is 0.25$-$0.9 mK, depending on the spectral range.

The IRAM 30m observations (project 123-22, PI Jim\'enez-Serra) were carried out in the period of February 1$-$18 2023. We used several frequency setups to cover different spectral ranges of the E0 (3 mm) and E1 (2 mm) bands of the Eight Mixer Receiver (EMIR) receiver. We used the Fast Fourier Transform Spectrometer (FTS), which provides a spectral resolution of 195 kHz. We slightly shifted in frequency each setup to identify possible contamination of lines from the image band. 
The new spectral ranges covered were 83.2$-$115.41 GHz, 132.28$-$140.39 GHz, and 142$-$173.81 GHz. The HPBW is $\sim$14$^{\prime\prime}$$-$29$^{\prime\prime}$. The spectra were exported from CLASS to \textsc{Madcuba}, and the reduction, including baseline subtraction, was carried out in \textsc{Madcuba}. We checked the line intensities of the new observations with those of previous data (e.g. \citealt{rodriguez-almeida2021a,rivilla2021a}), which are consistent within a 5$\%$ level. We then averaged the new data with those from the previous survey, weighting them by the noise of the spectra. 
The final spectra were smoothed to 615\,kHz (1.0$-$2.2 km s$^{-1}$ in the spectral range covered).
The final noise achieved is 0.5$-$2.5 mK at 3 mm, and 1.0$-$1.6 mK at 2 mm. For the spectral ranges not covered by these new observations, we used the data from our previous IRAM 30m survey (for details we refer to \citealt{rodriguez-almeida2021a,rivilla2022b,rivilla2022c}).

\begin{table*}
\centering
\caption{\ch{C2H5O2N} isomeric family. Relative energies, availability of microwave spectroscopy, dipole moments, and molecular abundances derived towards G+0.693 are indicated. The ``Microwave spectroscopy" column indicates the catalog entry when available, or refers to \textsc{Madcuba} for catalog entries generated from published spectroscopy.}
\begin{tabular}{l l c c c c }
\hline
Name    & Formula   & $\Delta E$    & Microwave   & $\mu$  & G+0.693  \\
        &           & kcal mol$^{-1}$ (K)  & spectroscopy   & Debye  & abundance ($\times$10$^{-10}$)  \\
\hline
N-methylcarbamic acid        & \ch{CH3NHCOOH}     &  0.0 (0.0) $^{a}$             &  --   &  2.5 $^{a}$   & --   \\  
$syn-$methyl carbamate   & \ch{CH3OC(O)NH2}   &  4.6 (2335)  $^{a}$      & JPL 75004 $^{e}$   & 2.3 $^{a}$     &  $<$ 2.5 \\ 
glycine (conformer I)   & \ch{NH2CH2COOH}   &  8.4 (4247) $^{a}$  &  CDMS 75511 $^{f}$ & 1.1 $^{i}$  &   $<$ 0.6  \\ 
glycine (conformer II)  & \ch{NH2CH2COOH} &  9.2  (4650) $^{b}$  &  CDMS 75512 $^{f}$ &  5.5 $^{i}$  &   $<$ 0.019 \\ 
$syn-$glycolamide      & \ch{NH2C(O)CH2OH}  &  9.7 (4866) $^{a}$        &  \textsc{Madcuba} $^{g}$  & 4.4 $^{a}$    &  0.55 \\ 
$anti-$glycolamide  & \ch{NH2C(O)CH2OH}  &  10.6 (5367) $^{c}$   &  \textsc{Madcuba} $^{g}$  & 2.6 $^{g}$    & $<$ 0.13 \\ 
N-(hydroxymethyl)formamide    & \ch{HOCH2NHCHO}               & 16.7  (8404)  $^{d}$      &  --  & 3.1 $^{d}$    &  -- \\
aminomethyl formate           & \ch{NH2CH2OCHO}            & 17.7  (8907)  $^{d}$      &  --   & 1.7 $^{d}$    & --  \\ 
aminohydroxyacetaldehyde      & \ch{NH2CH(OH)CHO}           & 27.6  (13889) $^{d}$     &  --  & 1.6 $^{d}$    & --  \\ 
2-aminoethene-1,1-diol        & \ch{NH2CHC(OH)2}           & 30.7  (15449)  $^{d}$      &  --  & 1.3 $^{d}$    & --  \\ 
O-acetylhydroxylamine        & \ch{CH3COONH2}     & 43.5 (21875) $^{a}$            &  --  &  1.9 $^{a}$     & --    \\ 
N-(Z)-hydroxyacetamide      & \ch{CH3C(O)NHOH}   & 44.7 (22474)  $^{a}$           &  \textsc{Madcuba} $^{h}$    & 3.3 $^{a,h}$      & $<$ 1.7  \\ 
N-hydroxyacetamidic acid   & \ch{CH3C(OH)NOH}   & 45.5 (22872)  $^{a}$             &  --    & 2.5 $^{a}$     & --   \\ 
\hline 
\end{tabular}
\label{tab:isomers}

{(a) Calculated at the CCSD(T)/aug-cc-pVTZ//MP2/aug-cc-pVTZ level by \citet{sanz-novo2019};}
{(b) Estimated using the B3LYP/6-311G(d,p) energy difference computed by \citet{lattelais2011};}
{(c) Estimated using the energy difference between syn- and anti-glycolamide calculated in \citet{maris2004} from relative intensity and dipole moment measurements;}
{(d) Calculated at the CCSD(T)/cc-pVQZ//B3LYP/cc-pVQZ level by \citet{lattelais2011};}
{(e) \citet{Groner2007}, \citet{ilyushin2006}, \citet{marstokk1999} and \citet{bakri2002};}
{(f) \citet{lovas1995} and \citet{ilyushin2005_glycine};}
{(g) \citet{sanz-novo2020} and \citet{maris2004};}
{(h) \citet{sanz-novo2022b};}
{(i) \citet{lovas1995}.}
\end{table*}

\section{Analysis and Results} 
\label{sec:analysis}

We list in Table \ref{tab:isomers} the \ch{C2H5O2N} isomeric family. We have searched for all the isomers for which rotational spectroscopy is available: $syn-$methyl carbamate (\ch{CH3OC(O)NH2}), conformers I and II of glycine\footnote{We note that there are up to thirteen conformers of glycine (\citealt{csaszar1992}), but we include in Table \ref{tab:isomers} only those two with lowest-energy, which are the only ones with rotational spectroscopy well characterized in the laboratory.} (\ch{NH2CH2COOH}), conformers $syn$ and $anti$ of glycolamide (\ch{NH2C(O)CH2OH}), and the $Z$-conformer of N-hydroxyacetamide (\ch{CH3C(O)NHOH}). 

The exquisite sensitivity of the new observations presented here has allowed to detect $syn$-glycolamide, which is the first \ch{C2H5O2N} isomer reported in the ISM. Glycolamide (or 2-hydroxyacetamide) has two conformers having {\it C}$_{s\rm}$ symmetry, with the hydroxyl (-OH) {\it syn} or {\it anti} with respect to the carbonyl group (-C=O). The {\it syn} conformer is lower in energy by $\sim$0.9 kcal mol$^{-1}$ (501 K; Table \ref{tab:isomers}), thanks to the formation of a stabilizing intramolecular hydrogen bond that also hinders a possible large amplitude motion of the $-$OH moiety.

For $syn$-glycolamide we have implemented the rotational spectroscopy from the laboratory work by \citet{sanz-novo2020} into the \textsc{Slim} (Spectral Line Identification and Modeling) tool within \textsc{Madcuba} (version 9.3.10, 04/05/2023). To search for the species, we generated with \textsc{Slim} a synthetic spectrum under the assumption of Local Thermodynamic Equilibrium (LTE) conditions. Figure \ref{fig-spectra} shows transitions of $syn$-glycolamide detected towards G+0.693 that are unblended or only slightly blended with transitions from other species. Spectroscopic information of these transitions is presented in Table \ref{tab:transitions}.
To ensure that the observed spectral features are due to $syn$-glycolamide, and to properly evaluate possible line contamination by other molecules, we have considered the emission from the $>$ 130 molecules already identified towards G+0.693 (e.g. \citealt{requena-torres_organic_2006,requena-torres_largest_2008,zeng2018,Rivilla2018,rivilla2019b,rivilla2020b,rivilla2021a,rivilla2021b,rivilla2022a,rivilla2022b,rivilla2022c,jimenez-serra2020,jimenez-serra2022,bizzocchi2020,rodriguez-almeida2021a,rodriguez-almeida2021b,zeng2021,colzi2022,alberton2023,sanandres2023,zeng2023}). 
We note that the rest of the $syn$-glycolamide transitions that fall within the spectral coverage of the survey are consistent with the observed spectra, but they are heavily blended with lines from other molecular species or too faint to be detected.

To derive the physical parameters of the emission, we used the \textsc{Slim-Autofit} tool of \textsc{Madcuba} that provides the best non-linear least-squares LTE fit to the data using the Levenberg-Marquardt algorithm. The parameters used in the LTE model are: molecular column density ($N$), excitation temperature ($T_{\rm ex}$), velocity ($v_\text{LSR}$) and full width at half maximum (FWHM) of the Gaussian line profiles. 
Since the molecular emission towards G+0.693 is extended and fills the telescope beams (\citealt{requena-torres_organic_2006,requena-torres_largest_2008,zeng2018,zeng2020}), no beam filling factor is used in our calculations.
The fit of the $syn$-glycolamide emission was performed by using the transitions shown in Figure \ref{fig-spectra}, and considering also the contribution of the other identified molecules. 
To allow the convergence of AUTOFIT, we fixed the value of  $T_{\rm ex}$ to 5 K, which was derived for the optically thin $^{13}$C isotopologue of the simplest amide, formamide, by \citet{zeng2023}. We also fixed the FWHM to 18 km s$^{-1}$, which reproduces well the profile of the most unblendend lines, and we left free $N$ and $v_\text{LSR}$. 
The best LTE fit is shown in Figure~\ref{fig-spectra}. 
We obtained a velocity of $v_\text{LSR}$ = 67.0 $\pm$ 1.0 km s$^{-1}$, which is in good agreement with the values found in G+0.693 for other molecular species, and in particular amides (\citealt{zeng2023}). The derived column density is (7.4 $\pm$ 0.7)$\times$10$^{12}$\,cm$^{-2}$, which translates into a molecular abundance with respect to molecular hydrogen of 5.5$\times$10$^{-11}$ (Table \ref{tab:isomers}), assuming $N_{\rm H_2}$ = 1.35$\times$10$^{23}$\,cm$^{-2}$ (\citealt{martin_tracing_2008}).

\begin{table*}
\centering
\tabcolsep 4.5pt
\caption{List of observed transitions of {\it syn}-glycolamide ({\it syn}-\ch{NH2C(O)CH2OH}). We provide the transitions frequencies, quantum numbers, base 10 logarithm of the integrated intensity at 300 K (log $I$), and the values of the upper levels of each transition ($E_{\rm u}$). The last column gives the information about the species whose transitions are partially blended with {\it syn}-glycolamide lines.}
\begin{tabular}{ c c c c l}
\hline
 Frequency & Transition  & log $I$ & $E_{\rm u}$  & Blending$^a$ \\
 (GHz) & ($J_{K_a,K_c}$)  &   (nm$^2$ MHz)  & (K) &   \\
\hline
 31.9403720   & 5$_{1,5}$ - 4$_{1,4}$  & -5.4635 & 4.9 &  unblended \\ 
 32.9834199  & 5$_{0,5}$ - 4$_{0,4}$  & -5.4258 & 4.8 & unblended \\ 
34.3346824    & 2$_{2,1}$ - 1$_{1,0}$  & -5.7880 & 2.3 &   partially blended with U  \\ 

 34.7930509    & 5$_{2,4}$ - 4$_{2,3}$  & -5.4485 & 6.3 &   partially blended with \ch{CH3COCH3}  \\ 
 34.9940974    & 5$_{1,5}$ - 4$_{0,4}$  & -5.3948 & 4.9 & partially blended with $^{13}$\ch{CH3CH2CN} \\ 
 35.4025207    & 5$_{3,3}$ - 4$_{3,2}$  & -5.5525 & 8.0 & partially blended with U \\ 
 35.5265875    & 2$_{2,0}$ - 1$_{1,1}$  & -5.7941 & 1.4 & blended with U \\ 
 36.8706638   &  6$_{0,6}$ - 5$_{1,5}$  & -5.2658 & 6.7 & unblended$^{*}$ \\ 
 36.8711421    & 5$_{2,3}$ - 4$_{2,2}$  & -5.3969 & 6.5 &  unblended$^{*}$ \\
 37.1038814    & 5$_{1,4}$ - 4$_{1,3}$  & -5.3359 &  5.7 &  unblended \\ 
 38.1071149   & 6$_{1,6}$ - 5$_{1,5}$  & -5.2286 & 6.7 &  partially blended with U  \\ 
 38.8813371    & 6$_{0,6}$ - 5$_{0,5}$  & -5.2068 & 6.7 &  unblended \\ 
 40.1177882    & 6$_{1,6}$ - 5$_{0,5}$  & -5.1784 & 6.7 &  unblended \\ 
41.5309974   & 6$_{2,5}$ - 5$_{2,4}$  & -5.1944 & 8.3 & unblended \\ 
43.4830710    & 7$_{0,7}$ - 6$_{1,6}$ &  -5.0310 & 8.8 & partially blended with $anti-$ and $syn-$\ch{CH2CHCHNH} \\
 44.2047398   & 7$_{1,7}$ - 6$_{1,6}$  & -5.0327 & 8.9 & partially blended with $agG-$(\ch{CH2OH})$_2$ \\ 
 44.7195221   & 7$_{0,7}$ - 6$_{0,6}$  & -5.0209 & 8.8 & partially blended with $agG-$(\ch{CH2OH})$_2$ \\ 
88.0821975    & 7$_{3,4}$ - 6$_{2,5}$  & -4.8243 & 12.5 & unblended \\ 
 97.6105934    & 5$_{5,1}$ - 4$_{4,0}$  & -4.4215 &  13.3  & unblended$^{*}$ \\ 
 97.6111554     & 5$_{5,0}$ - 4$_{4,1}$  & -4.4214 & 13.3  & unblended$^{*}$ \\  
\hline 
\end{tabular}
\label{tab:transitions}
{\\ (a) ``U" refers to blendings with emission from an unknown (not identified) species; transitions with the $^{*}$ symbol are not blended with emission from other species, but (auto)blended with another transition of {\it syn}-glycolamide.}
\end{table*}

We also searched for the other \ch{C2H5O2N} isomers for which spectroscopy is available (see Table \ref{tab:isomers}): the higher-energy $anti$-conformer of glycolamide, the conformers I and II of glycine (\ch{NH2CH2COOH}), $syn-$methyl carbamate (\ch{CH3OC(O)NH2}), and the $Z$ isomer of N-hydroxyacetamide (\ch{CH3C(O)NHOH}). None of them is detected, so we computed the upper limits of their column densities using the catalog entries and spectroscopic works presented in Table \ref{tab:isomers}. The detailed description of the calculations and the specific transitions used are presented in Appendix \ref{app:upper-limits}, and the resulting upper limits are listed in Table \ref{tab:isomers}.

\section{Discussion}
\label{sec:discussion}

\subsection{The first \ch{C2H5O2N} isomer detected in the ISM}

In spite of the many observational efforts done in the past, mainly targeting glycine, no \ch{C2H5O2N} isomer was reported in the ISM before the detection of glycolamide presented in this work. We discuss in this section why glycolamide is easier to detect than other isomers, and, in particular, the long-sought glycine.

The detectability of a molecular species through rotational spectroscopy depends, obviously, on the molecular abundance, but also on the strength of the dipole moment ($\mu$). The larger the dipole moment the brighter the line intensities, and thus the easier to detect the molecule. We summarize in Table \ref{tab:isomers} the dipole moments of the different \ch{C2H5O2N} isomers, computed from the experimental and theoretical works cited there. The highest dipole moments are those of the higher-energy conformer of glycine (conformer II) and $syn$-glycolamide (5.5 and 4.4 Debye, respectively). 

Besides the dipole moment, the relative energies between isomers have been used  traditionally to predict if a molecule might be present in the ISM.
\citet{lattelais2011} proposed that the isomers with higher abundances are those most thermodynamically stable. This empirical rule was called the Minimum Energy Principe (MEP).
However, the broad variety of new isomers detected in the last years has cast serious doubts about the applicability of the MEP. 
It still holds for stereoisomers (or spatial isomers),\footnote{Isomers that possess identical constitution, but differ in the three-dimensional 
orientations of their atoms.} such as the $anti$ and $gauche$ conformers of ethyl formate (\ch{CH3CH2(O)CHO}; \citealt{tercero_discovery_2013,rivilla_chemical_2017}), or the $Z$ and $E$ isomers of cyanomethanimine (\ch{HNCHCN}).
Their relative isomeric ratios follow the value expected according to the thermodynamic equilibrium, which scales with exp(-$\Delta E$/$T_{\rm k}$), where $\Delta E$ is the energy difference between the conformers and $T_{\rm k}$ the kinetic temperature of the gas. \citet{garciadelaconcepcion2021} recently demonstrated, using theoretical calculations, that this equilibrium is indeed possible in the conditions of the ISM thanks to tunneling effects that allow the isomerization.
  
However, it is well known that MEP fails to predict the relative abundances of many constitutional (or structural) isomers,\footnote{Molecular species with the same number of atoms of each element, but arranged with different bonds between them.} e.g. the \ch{C3H2O} or \ch{C2H4O2} and isomeric families (see e.g. \citealt{shingledecker2019} and \citealt{mininni2020}, respectively).

\begin{figure}
\begin{center}
\includegraphics[width=8.4cm]{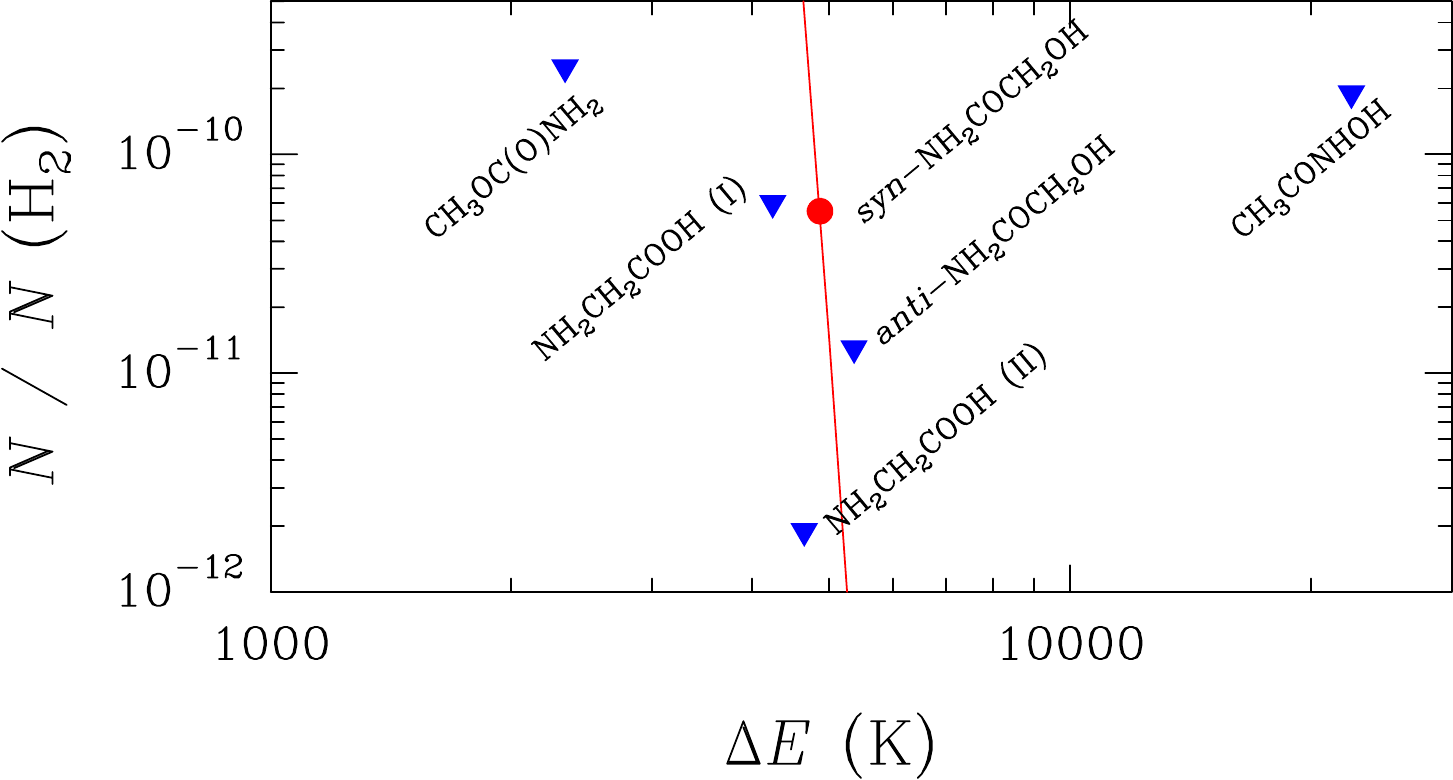}
\vspace{-5mm}
\end{center}
\caption{Molecular abundances with respect to H$_2$ of \ch{C2H5O2N} isomers as a function of its relative energy (see Table \ref{tab:isomers}). The red dot corresponds to the  detection of $syn$-glycolamide, while the blue triangles indicate the upper limits derived for the other isomers. The solid red line indicates the expected molecular abundance of the \ch{C2H5O2N} isomeric family assuming the thermodynamic population at $T_{\rm k}$ = 100 K, and using the abundance of $syn$-glycolamide as reference.}
\label{fig-isomers}
\end{figure}

The analysis of the \ch{C2H5O2N} isomers presented in this work confirms that the MEP does not hold for this family either. 
We show in Figure \ref{fig-isomers} the molecular abundances with respect to H$_{2}$ of the 
\ch{C2H5O2N} isomers as a function of its relative energies, listed in  Table \ref{tab:isomers}.
The most stable isomer with available spectroscopy is $syn-$methyl carbamate, followed by the both conformers of glycine, none of them detected towards G+0.693.  
Glycolamide is the next isomer in terms of relative energy. The red line in Figure \ref{fig-isomers} indicates the expected molecular abundance of the \ch{C2H5O2N} isomers as a function of their relative energy, in case the thermodynamic equilibrium were governing the isomeric relative abundances. We have assumed a kinetic temperature of $T_{\rm k} =$ 100 K for G+0.693 (\citealt{zeng2018}), and used the relative energy differences in Table \ref{tab:isomers}. From thermodynamics, N-methylcarbamic acid and both glycine conformers should have much higher abundances than glycolamide, and also higher than the upper limits derived for them (blue triangles in Figure \ref{fig-isomers}). Therefore, the abundances of all the \ch{C2H5O2N} isomers cannot be explained in terms of thermodynamical equilibrium, and hence they need to be described by chemical kinetics, i.e., by different formation and destruction routes for each isomer. 

In the case of the most-searched isomer, glycine, its non-detection might not be surprising from an astrochemical point of view. While chemical compounds with functional groups such as \ch{-CH3}, \ch{-CH2CH3}, \ch{-C+N}, \ch{-CH2OH}, \ch{-HCO}, or \ch{-NH2}, are common,  interstellar molecules with the carboxyl group ($-$COOH) are remarkably scarce. So far, only the simplest representatives, formic acid (HCOOH) and acetic (\ch{CH3COOH}), have been reported; and only very recently also carbonic acid (HOCOOH) has been also detected (\citealt{sanz-novo2023}).

Glycine (\ch{NH2CH2COOH}) can be seen as the amine derivative of acetic acid (\ch{CH3COOH}), in which one H atom has been replaced by the amine group (\ch{NH2}). Thus, we can use other H/\ch{NH2} pairs already detected towards G+0.693 to infer the expected abundance of glycine: glycolaldehyde (\ch{HCOCH2OH}) and glycolamide (\ch{NH2COCH2OH}), and ethanol (\ch{CH3CH2OH}) and ethanolamine (\ch{NH2CH2CH2OH}). The glycolaldehyde/glycolamide ratio is $\sim$15 (\citealt{rivilla2022a}, and this work), while the ethanol/ethanolamine ratio is $\sim$40 (\citealt{rivilla2021a,rodriguez-almeida2021a}). If we assume these ratios for the acetic acid/glycine pair, and using the abundance of acetic acid reported by \citet{sanz-novo2023} of 3.1$\times$10$^{-10}$, the abundance of glycine would be in the range $\sim$(0.8$-$2)$\times$10$^{-11}$. These values are a factor of 3$-$8 below the upper limit derived in this work for the low-energy conformer I (Table \ref{tab:isomers}). As discussed in more detail in Appendix \ref{app:upper-limits}, if both glycine conformers were in thermodynamic equilibrium, the measured upper limit derived for the high-energy conformer (II) would translate into an upper limit for the low-energy conformer of $\sim$10$^{-10}$, very similar to that derived using the conformer I itself. Hence, it is clear that the abundance of glycine towards G+0.693 is expected to be below 10$^{-10}$, and surely lower than that derived for its isomer glycolamide (see Figure \ref{fig-isomers}).

Therefore, the chemistry that takes places in the ISM might be able to form glycine, but likely its low expected abundance makes very challenging to detect it with present-day radiotelescopes at reasonable integration times. The detection of glycolamide indicates that it can be efficiently formed under interstellar conditions. In next section we will discuss several chemical pathways that can form glycolamide in the ISM.

\begin{figure*}
\begin{center}
\includegraphics[width=19cm]{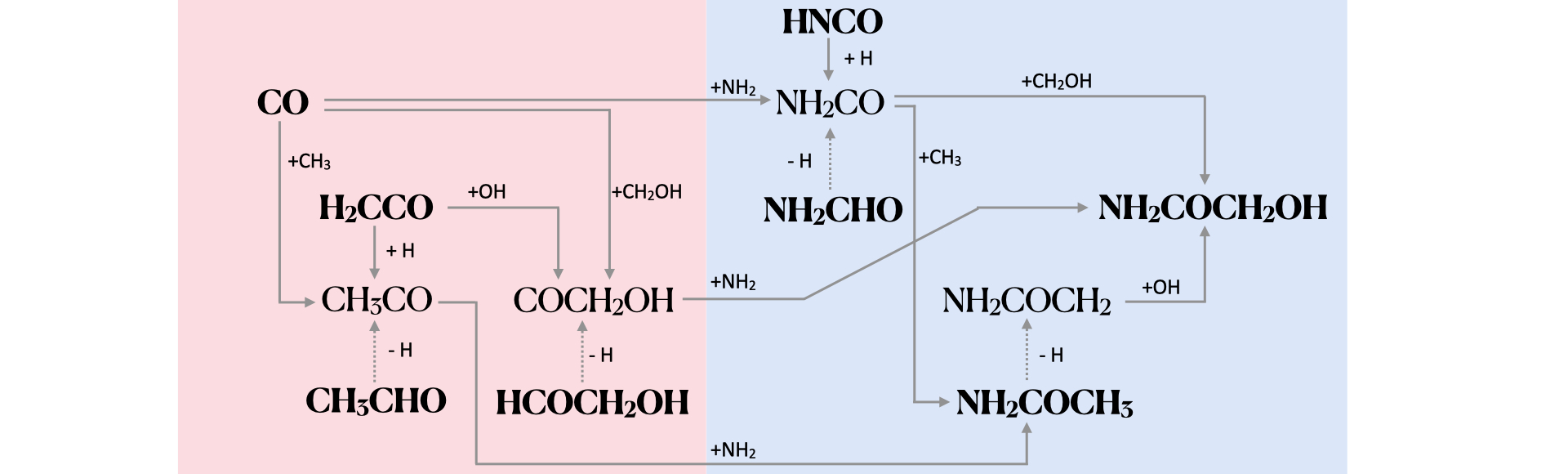}
\vspace{-2mm}
\end{center}
\caption{Proposed chemical network for the formation of glycolamide in the ISM. The left red area includes oxygen-bearing species, while the blue right panel includes nitrogen- and oxygen-bearing species. Closed-shell species are labeled in boldface, as opposed to radicals.}
\label{fig-chem-diagram}
\end{figure*}

\subsection{Formation of glycolamide}
\label{sec:formation-glycolamide}

The chemistry of G+0.693 is thought to be dominated by large-scale shocks  (\citealt{martin-pintado2001,martin_tracing_2008}) produced by a cloud-cloud collision (\citealt{zeng2020}). These shocks are responsible for the sputtering of dust grains, releasing many molecules formed on the grain surfaces into the gas phase (\citealt{caselli1997,jimenez-serra2008}). Therefore, surface chemistry is expected to play a major role in G+0.693. For this reason, and because no viable gas-phase route of glycolamide has been proposed in the literature, from here on we will consider only grain-surface reactions.

Glycolamide is the largest molecule containing an amide functional group ($-$NH-C(=O)$-$) detected in the ISM. It is an $\alpha$-substitute of acetamide (\ch{NH2C(O)CH3}), in which one hydrogen atom H of the methyl group is replaced by the hydroxyl group $-$OH. Based on this similarity, both amides might have similar formation routes. The chemistry of acetamide in G+0.693 has been recently discussed by \citet{zeng2023}. Chemical models (\citealt{belloche2017,belloche2019,garrod2022}) have suggested that the formation of acetamide involves reactions between simple radicals (\ch{CH3} and \ch{NH2}) with larger radicals (\ch{NH2CO} and \ch{CH3CO}) that can be synthesized on grains through H-addition and/or H-abstraction reactions of species that are abundant in the ISM, and in G+0.693 in particular (\ch{HNCO}, \ch{NH2CHO}, \ch{H2CCO}, and \ch{CH3CHO}); and/or through the combination of CO with \ch{NH2} (see the experiments of \citealt{ligterink2018}) and \ch{CH3}, respectively (see Figure \ref{fig-chem-diagram}).

Analogous pathways involving radical-radical reactions, depicted in Figure \ref{fig-chem-diagram}, are promising formation routes for glycolamide. \citet{sanz-novo2020} proposed that it can be formed from the reaction between \ch{NH2} and the \ch{COCH2OH} radical. The latter can result from the H-abstraction on the formyl site of glycolaldehyde (\ch{HCOCH2OH}), the reaction between CO with the hydroxymethyl radical (\ch{CH2OH}), and/or the combination of the OH radical with ketene (\ch{H2CCO}).
\citet{haupa2020} suggested that glycolamide can also be formed through the radical-radical reaction between \ch{OH} and the 2-amino-2-oxoethyl radical (\ch{NH2COCH2}; Figure \ref{fig-chem-diagram}), which is formed by the H-abstraction on the methyl site of acetamide according to their experiments.
A third possible route was proposed by \citet{garrod_complex_2008}: the reaction of \ch{NH2CO} (a precursor of acetamide, as mentioned above, which is formed by hydrogen
abstraction from formamide), and \ch{CH2OH} (Figure \ref{fig-chem-diagram}). 
 
All of the possible closed-shell precursors are relatively abundant towards G+0.693. Their molecular abundances with respect to \ch{H2}, in units of $10^{-10}$, are: 266 (\ch{HNCO}; \citealt{zeng2018}); 37 (\ch{CH3CHO}; \citealt{Sanz-Novo2022}); 21 (\ch{H2CCO}; \citealt{requena-torres_largest_2008}\footnote{The molecular abundance reported by \citet{requena-torres_largest_2008} used an \ch{H2} column density for G+0.693 of 4.1$\times$10$^{22}$ cm$^{-2}$, so we have recalculated it using the same value we have used for the other species, which is 1.35$\times$10$^{22}$ cm$^{-2}$ (\citealt{martin_tracing_2008}).}); 18.5 (\ch{NH2CHO}; \citealt{zeng2023}); 8.5 (\ch{NH2COCH3}; \citealt{zeng2023}); and 6.9 (\ch{HCOCH2OH}; \citealt{rivilla2022a}); which are higher than the abundance of $syn$-glycolamide by factors of $\sim$15-500. Hence, all of them are viable precursors. 

\begin{figure*}
\begin{center}
\includegraphics[width=13cm]{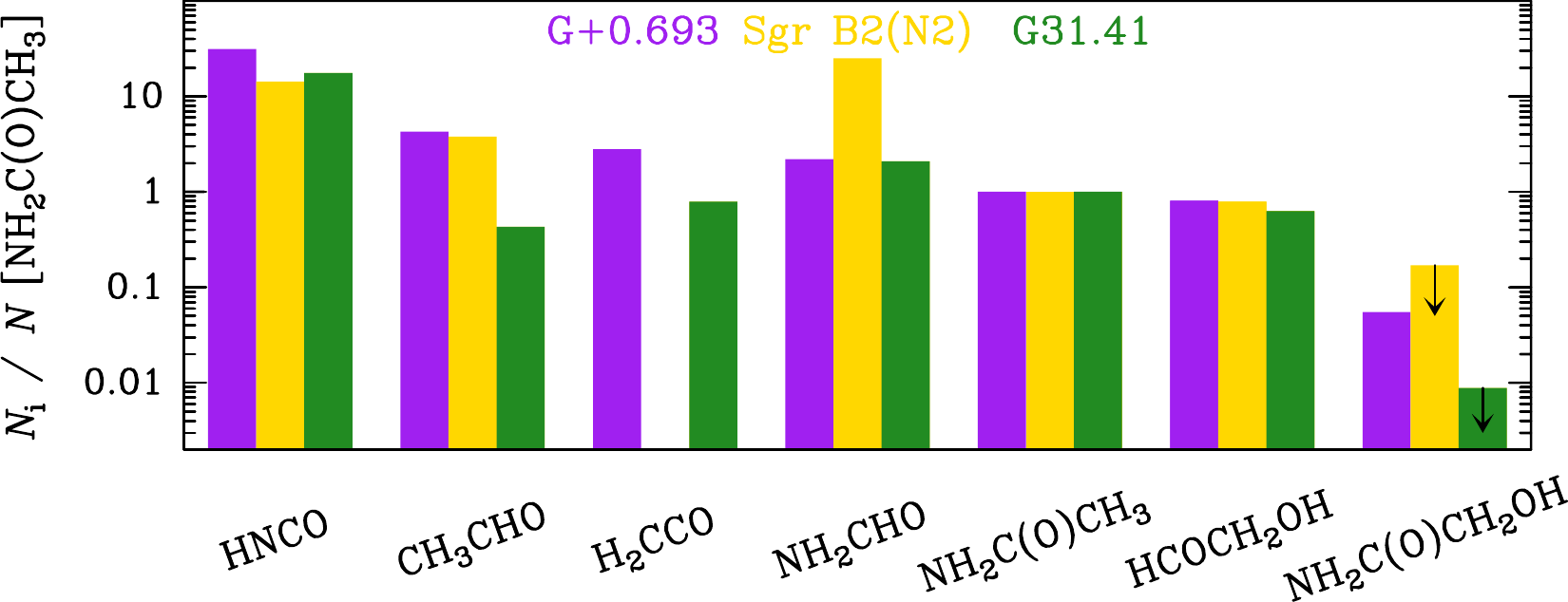}
\vspace{-5mm}
\end{center}
\caption{Comparison of the relative molecular abundances, compared to acetamide, of the proposed precursors of glycolamide, derived towards G+0.693 (purple) and the hot cores Sgr B2(N2) (yellow) and G31.41+0.31 (green). Upper limits of glycolamide are indicated with arrows.}
\label{fig:histo-abundances}
\end{figure*}

Besides G+0.693, abundance upper limits of glycolamide have been derived towards the hot molecular cores Sgr B2(N2), located $\sim$60$^{\prime\prime}$ southwest from G+0.693, and G31.41+0.41 (hereafter G31), located in the Galactic disk (\citealt{sanz-novo2020} and \citealt{colzi2021}, respectively).
We have compared in Figure \ref{fig:histo-abundances} the column densities (normalized to acetamide, \ch{NH2C(O)CH3}) derived in the three sources. 
We used the molecular abundances from \citet{zeng2018}, \citet{Sanz-Novo2022}, \citet{rivilla2022a}, and \citet{zeng2023} for G+0.693; \citet{mininni2020}, \citet{colzi2021} and private communication for G31; and \citet{belloche2017} and \citet{sanz-novo2020} for Sgr B2(N2).
The proposed precursors of glycolamide have similar molecular ratios in  Sgr B2(N2) and G+0.693, within factors of 1.0$-$2.2, with the only exception of \ch{NH2CHO}, which is an order of magnitude more abundant in Sgr B2(N2) (\citealt{belloche2017}). This might imply that \ch{NH2CHO} is not the dominant precursor of \ch{NH2C(O)CH2OH}, because in that case, the formation of glycolamide would be enhanced by the same factor toward Sgr B2(N2) as compared to the other sources.

The molecular abundance ratios in G31 are very similar to those in G+0.693 within factors $\sim$1.0$-$3.5, with the exceptions of \ch{CH3CHO} and \ch{NH2C(O)CH2OH}, which are lower by factors of $\sim$10 and $>$6.3, respectively. 
This might indicate that acetaldehyde and glycolamide are chemically linked. 
However, we note that firm conclusions cannot be drawn based only on three astronomical sources, and that the detection of glycolamide in more astronomical environments (molecular clouds, hot cores/corinos, protostellar shocks) is needed to shed light on the formation processes of glycolamide. 

Since radical-radical reactions on the surface of dust grains are generally barrierless, the bottleneck of the proposed pathways is the formation of the larger radicals shown in Figure \ref{fig-chem-diagram}:  
\ch{CH3CO}, \ch{COCH2OH}, \ch{NH2CO}, and \ch{NH2COCH2}. 
The detection of glycolamide towards G+0.693 suggests that the production of these radicals is more efficient than in Sgr B2(N2) and G31. 
These radicals can be formed from abundant closed-shell species by H$-$abstractions, and H$-$, OH$-$, \ch{CH3}$-$ and \ch{NH2}$-$ additions (Figure \ref{fig-chem-diagram}). 
All these mechanisms can be enhanced in the presence of a ultraviolet field induced by cosmic rays (\citealt{garrod2022}). 
Indeed, a high cosmic-ray ionisation rate (CRIR) has been derived towards G+0.693 ($\sim$10$^{-15}-$10$^{-14}$ s$^{-1}$, \citealt{rivilla2022b}), in good agreement with those measured in its hosting Sgr B2 region (\citealt{yusef-zadeh2007,yusef-zadeh2016}), and across the whole Galactic Center (\citealt{goto_cosmic_2013}). Although Sgr B2(N2) is located close to G+0.693, the extremely high column density of \ch{H2} the hot core ($\sim$10$^{26}$ cm$^{-2}$, \citealt{sanchez-monge2017}; much higher than that of G+0.693 of $\sim$10$^{23}$ cm$^{-2}$) is expected to drastically attenuate the CRIR (\citealt{padovani2013}). Similarly, G31 is exposed to a much lower cosmic-ray flux since it is located in the Galactic disk, where the CRIR is 2$-$3 orders of magnitude lower than in the Galactic Center, which is further attenuated due to the high \ch{H2} column density of the molecular core ($\sim$10$^{25}$ cm$^{-2}$; \citealt{mininni2020}).
In summary, the formation of complex molecules such as \ch{NH2COCH2OH} on the grain surfaces through filling reactions is expected to be boosted in the presence of high CRIR, like that present in G+0.693.

\section{Summary and Conclusions}

We have searched for members of the isomeric family of glycine (\ch{C2H5O2N}) towards the G+0.693-0.027 molecular cloud located in the Galactic Center, using new ultradeep observations carried out with the Yebes 40m and IRAM 30m telescopes with improved sensitivity at sub-mK levels. We present the first detection in the interstellar medium of a \ch{C2H5O2N} isomer: the $syn$ conformer of glycolamide (\ch{NH2C(O)CH2OH}). 
We have derived a column density of (7.4 $\pm$ 0.7)$\times$10$^{12}$\,cm$^{-2}$, which translates into a molecular abundance with respect to molecular hydrogen of 5.5$\times$10$^{-10}$. 
We also report upper limits of the molecular abundances of five other isomers: the $anti$ conformer of glycolamide, the conformers I and II of glycine (\ch{NH2CH2COOH}), $syn-$methyl carbamate (\ch{CH3OC(O)NH2}), and N-(Z)-hydroxyacetamide (\ch{CH3C(O)NHOH}). 
The upper limit for the low-energy conformer of glycine is $<$ 0.6$\times$10$^{-10}$, which indicates that it is less abundant than $syn$-glycolamide. The relative abundances of the \ch{C2H5O2N} isomers cannot be described in terms of thermodynamics, and thus they are due to different chemical pathways in the ISM, which favour the formation of glycolamide over other isomers.
We discuss different routes to produce glycolamide on the surface of dust grains, based on reactions between simple radicals (\ch{OH}, \ch{NH2}, \ch{CH3} and \ch{CH2OH}) and larger radicals generated from abundant precursors already detected in the cloud (CO, \ch{HNCO}, \ch{CH3CHO}, \ch{NH2CHO}, \ch{H2CCO}, \ch{HCOCH2OH} or \ch{NH2COCH3}) after H$-$, OH$-$ and \ch{NH2}$-$additions, and/or H$-$abstractions. The formation of these radicals is expected to be enhanced in the presence of the intense cosmic-ray secondary ultraviolet field likely present in G+0.693, providing a natural explanation for the detection of glycolamide, and opening the window for the detection of equally or even more complex species.

\software{Madrid Data Cube Analysis (\textsc{Madcuba}) on ImageJ is a software developed at the Center of Astrobiology (CAB) in Madrid; \url{https://cab.inta-csic.es/madcuba/}; \citep{martin2019}; GILDAS,  \url{https://www.iram.fr/IRAMFR/GILDAS}}.

\begin{acknowledgments}
We acknowledge the two anonymous reviewers for their careful reading of the manuscript and their useful comments.
We are grateful to the IRAM 30m and Yebes 40m telescopes staff for their help during the different observing runs.
The 40m radio telescope at Yebes Observatory is operated by the Spanish Geographic Institute (IGN, Ministerio de Transportes, Movilidad y Agenda Urbana).
IRAM is supported by INSU/CNRS (France), MPG (Germany) and IGN (Spain). 
V.M.R. acknowledges support from the project RYC2020-029387-I funded by MCIN/AEI/10.13039/501100011033.
M.S.N. thanks the financial funding from the European Union - 
NextGenerationEU, Ministerio de Universidades and the University of Valladolid under a postdoctoral Margarita Salas Grant, as well as financial support from the Spanish Ministerio de Ciencia e Innovación (PID2020-117742GB-I00).
I.J.-S., J.M.-P., L.C., A.M and A.M.-H. acknowledge funding from grants No. PID2019-105552RB-C41 and MDM-2017-0737 (Unidad de Excelencia Mar\'ia de Maeztu-Centro de Astrobiolog\'ia, INTA-CSIC) by the Spanish Ministry of Science and Innovation/State Agency of Research MCIN/AEI/10.13039/501100011033 and by ``ERDF A way of making Europe".
D.S.A acknowledges the funds provided by the Consejo Superior de Investigaciones Cient{\'i}ficas (CSIC) and the Centro de Astrobiolog{\'i}a (CAB) through the project 20225AT015 (Proyectos intramurales especiales del CSIC). 
P.dV. and B.T. thank the support from the Spanish Ministerio de Ciencia e Innovaci\'on (MICIU) through project PID2019-107115GB-C21. B.T. also thanks the Spanish MICIU for funding support from grant PID2019-106235GB-I00.
J.L.A. acknowledges the fundings from Ministerio de Ciencia e Innovaci\'on (PID2019-111396GB-I00) and European Research Council  ERC-2013-SyG, Grant Agreement n. 610256 NANOCOSMOS.

\end{acknowledgments}

\bibliography{glycolamide}{}
\bibliographystyle{aasjournal}

\newpage
\appendix
\twocolumngrid
\restartappendixnumbering

\section{Upper limits of non-detected \ch{C2H5O2N} isomers}
\label{app:upper-limits}

\begin{figure}
\begin{center}
\includegraphics[width=9cm]{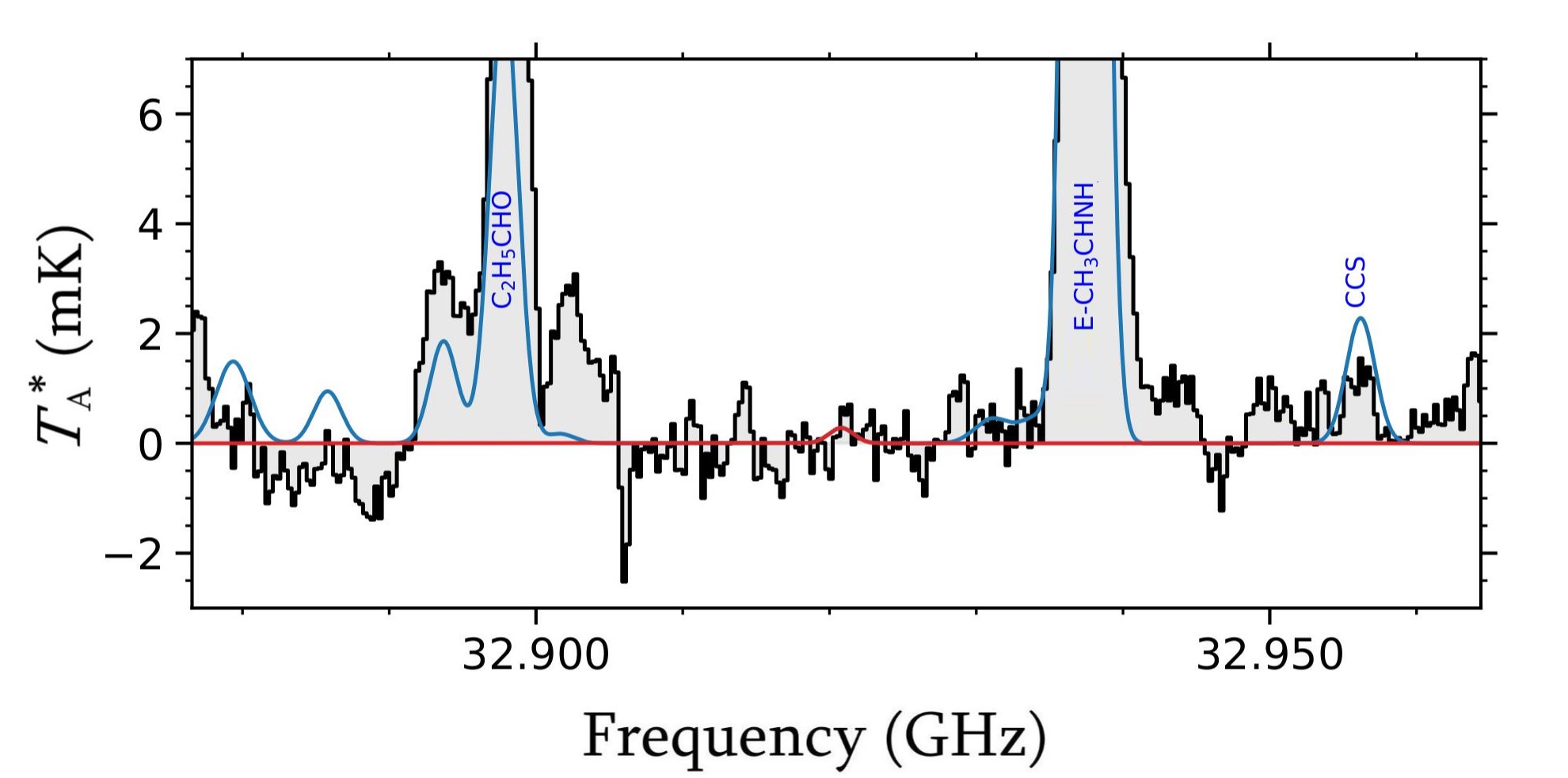}
\end{center}
\vspace{-5mm}
\caption{Transition of $anti$-glycolamide ($anti$-\ch{NH2C(O)CH2OH}) used to derive its column density upper limit. The observed spectra are shown as a grey histogram, and the LTE synthetic model using the derived upper limit is shown with a red curve.}
\label{fig:anti}
\end{figure}

We have also searched for other \ch{C2H5O2N} isomers for which rotational spectroscopy is available, which are not detected. We have thus derived 3$\sigma$ upper limits for their molecular abundances, where $\sigma$ is the root mean square noise of the spectra. We used  the brightest transitions predicted by the LTE model that appear completely unblended, which are listed in Table \ref{tab:limits}. For consistency, we used the same $T_{\rm ex}$, $v_\text{LSR}$, and FWHM values of the detected $syn$ conformer of glycolamide. 

For the high-energy $anti$ conformer of glycolamide, we imported into \textsc{Slim} the spectroscopic parameters reported in Table 2 of \citet{sanz-novo2020}, which include the experimental measurements and dipole moment calculations from \citet{maris2004}.
We used the hyperfine transitions 5$_{0,5}-$4$_{0,4}$ (at 32.9205744 and 32.9208273 GHz), shown in Figure \ref{fig:anti}. 
We obtained $N<$ 1.7$\times$10$^{12}$\,cm$^{-2}$, which translates into a molecular abundance compared to \ch{H2} of 0.13$\times$10$^{-10}$ (Table \ref{tab:isomers}). The derived $syn/anti$ ratio is $>$4.4, which is consistent with the one predicted if both conformers follow thermodynamic equilibrium, as observed for other species in the ISM (e.g. \citealt{rivilla_chemical_2017,rivilla2019b,garciadelaconcepcion2021}). Using their relative energy difference ($\Delta E$= 501 K, see Table \ref{tab:isomers}), and using a kinetic temperature of $T_{\rm k} =$ 100 K for G+0.693 (\citealt{zeng2018}), their expected relative ratio would be $syn/anti>$ 150, which indicates that the detection of the $anti$ conformer would be challenging.

\begin{table*}
\centering
\tabcolsep 4.5pt
\caption{Transitions used to compute the molecular abundance upper limits for the \ch{C2H5O2N} isomers not detected towards G+0.693. We provide the quantum numbers, transitions frequencies, base 10 logarithm of the integrated intensity at 300 K (log $I$), and the values of the upper levels for each transitions ($E_{\rm u}$). }
\begin{tabular}{c c c c c c}
\hline
Species & Formula &  Transition &  Frequency & log $I$ & $E_{\rm u}$  \\
 & &  & (GHz)    &   (nm$^2$ MHz)  & (K)   \\ 
\hline
$anti$-glycolamide    & \ch{NH2C(O)CH2OH}   & 5$_{0,5,6}$ - 4$_{0,4,5}$$^{(a)}$ & 32.9205744 & -5.9032 & 4.8 \\ 
                      &                     & 5$_{0,5,5}$ - 4$_{0,4,4}$  & 32.9208273  &  -5.9935 & 4.8 \\ 
glycine (conformer I)     & \ch{NH2CH2COOH}   & 5$_{1,5}$ - 4$_{1,4}$$^{(b)}$ & 31.1583057 & -6.5064 & 4.8 \\ 
                          &    & 5$_{0,4}$ - 4$_{0,4}$ & 32.2028617 & -6.4669 & 4.7 \\ 
glycine (conformer II)    & \ch{NH2CH2COOH}   & 6$_{2,5}$ - 5$_{2,4}$$^{(b)}$ & 41.9020415 & -4.6541 & 8.3 \\ 
$syn-$methyl carbamate           & \ch{CH3OC(O)NH2}  & 2$_{2,0,2}$ - 1$_{1,1,1}$ A$^{(c,d)}$ & 36.7150325 & -7.1378 & 2.4 \\
                                &   & 2$_{2,0,2}$ - 1$_{1,1,2}$ A & 36.7156362 & -7.6149 & 2.4 \\ 
                                &   & 2$_{2,0,1}$ - 1$_{1,1,1}$ A & 36.7164348 & -7.6150 & 2.4 \\ 
                                &   & 2$_{2,0,3}$ - 1$_{1,1,2}$ A & 36.7165377 & -6.8668 & 2.4 \\ 
                                &   & 2$_{2,0,1}$ - 1$_{1,1,0}$ A & 36.7179444 & -7.4900 & 2.4 \\ 
N-(Z)-hydroxyacetamide    & \ch{CH3C(O)NHOH}  & 6$_{1,6,7}$ - 5$_{0,5,6}$ A$^{(d)}$  & 40.8305064 & -6.6448 & 6.9\\ 
\hline 
\end{tabular}
\label{tab:limits}
\tablecomments{$^{(a)}$ For $anti$-glycolamide and N-(Z)-hydroxyacetamide the Hamiltonian was set up in the coupled basis set $I$ + $J$ = $F$, so the energy levels involved in each transition are labeled with the quantum numbers $J$, $K$$_{a}$, $K$$_{c}$, and $F$.
{$^{(b)}$ For the glycine conformers, the quantum numbers are $J$, $K_a$ and $K_c$}.
$^{(c)}$ For $syn-$methyl carbamate, an internal rotor Hamiltonian was used and the hyperfine structure was also considered. 
$^{(d)}$ The $A$ label refers to the $A$-symmetry state of $syn-$methyl carbamate and N-(Z)-hydroxyacetamide, respectively, which is due in both cases to the presence of a methyl internal rotation motion.}
\end{table*}

\begin{figure}
\begin{center}
\includegraphics[width=9cm]{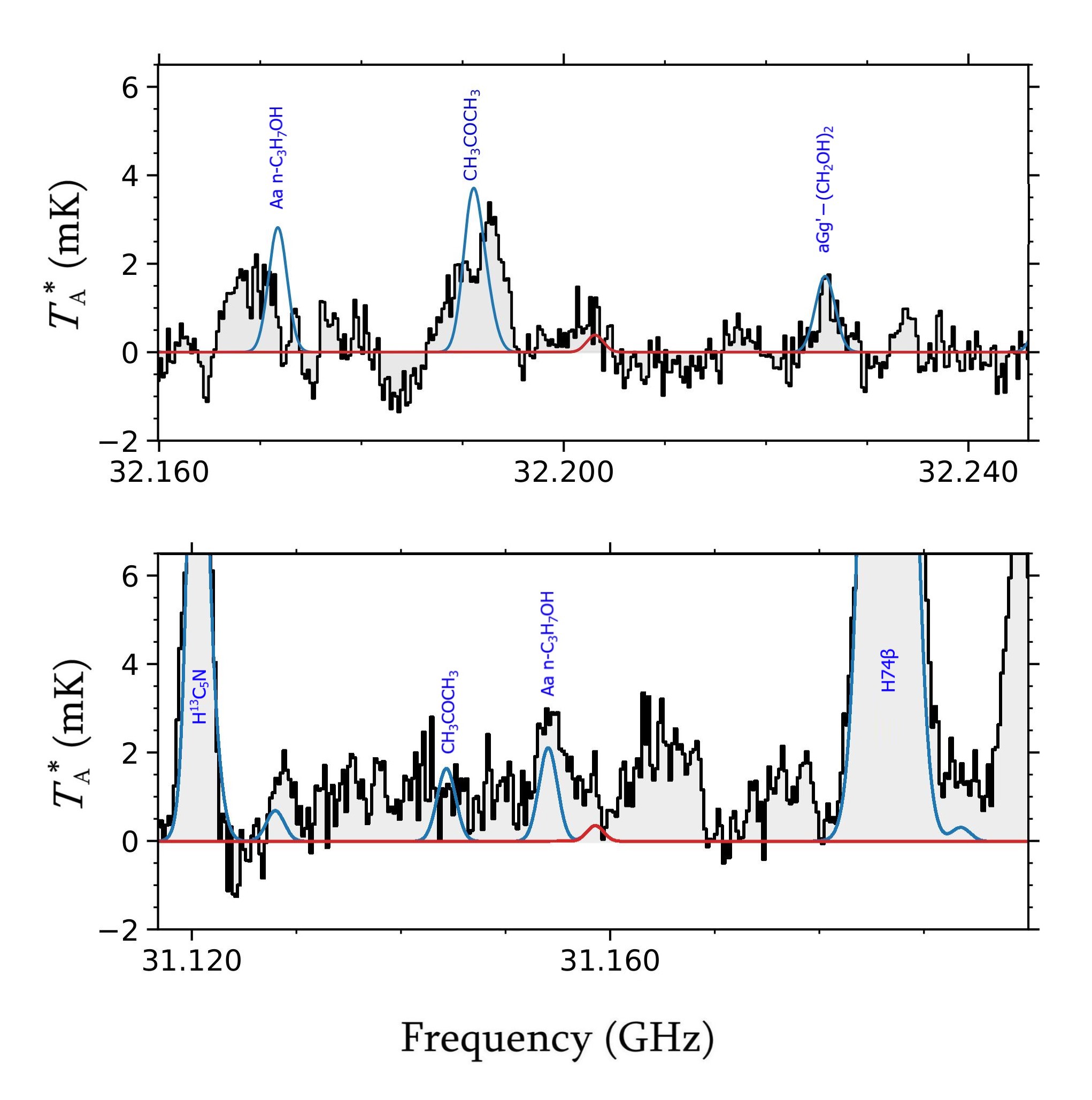}
\end{center}
\vspace{-5mm}
\caption{Transitions of the conformer I of glycine (\ch{NH2CH2COOH})  used to derive its column density upper limit. The observed spectra are shown as a grey histogram, and the LTE synthetic model using the derived upper limit is shown with a red curve.}
\label{fig:glycineI}
\end{figure}

\begin{figure}
\begin{center}
\includegraphics[width=9cm]{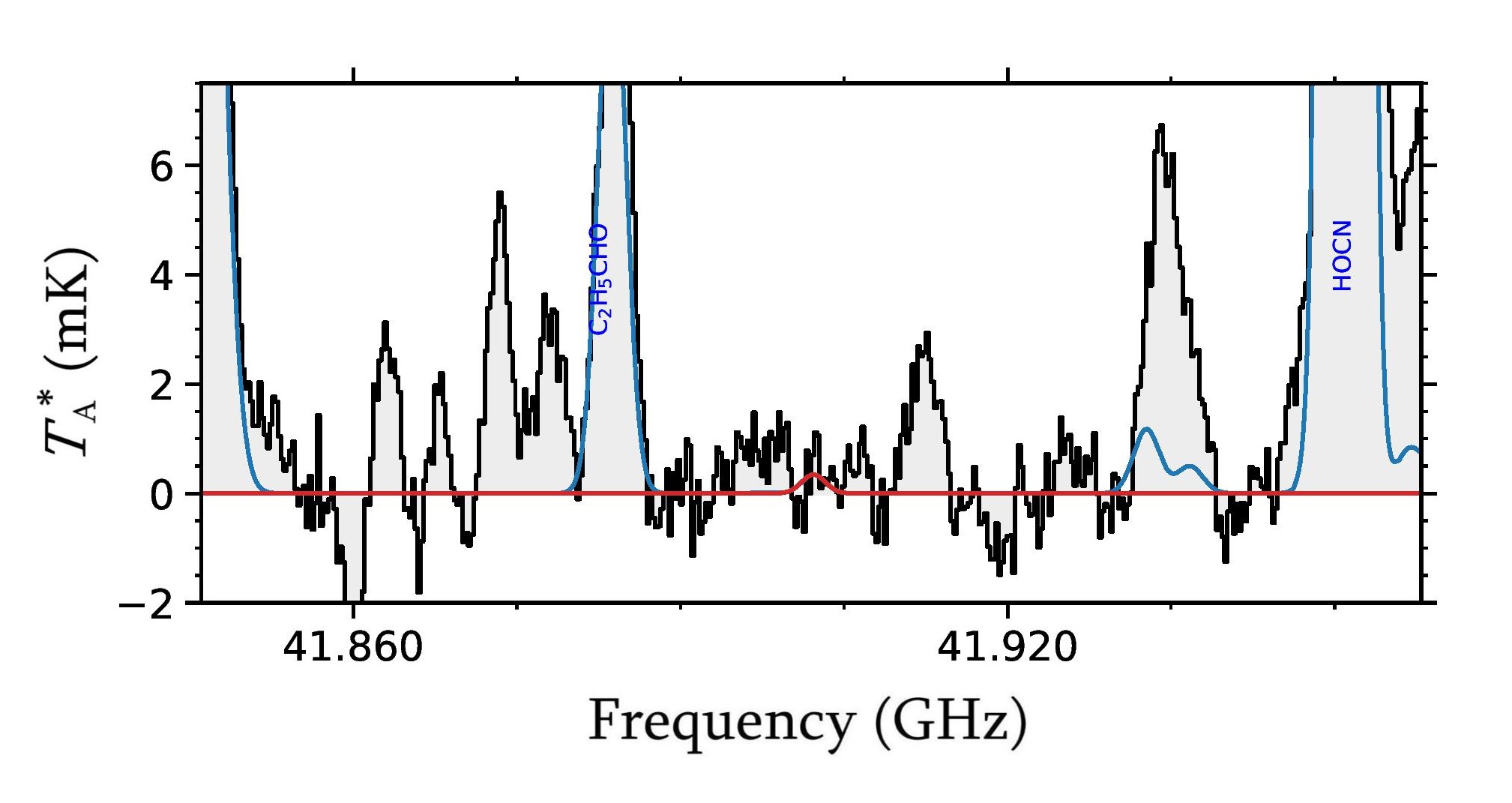}
\end{center}
\vspace{-5mm}
\caption{Transitions of the conformer II of glycine (\ch{NH2CH2COOH})  used to derive its column density upper limit. The observed spectra are shown as a grey histogram, and the LTE synthetic model using the derived upper limit is shown with a red curve.}
\label{fig:glycineII}
\end{figure}

For the conformers I and II of glycine (\ch{NH2CH2COOH}) we used the Cologne Database for Molecular Spectroscopy (CDMS, \citealt{endres2016}) entries 75511 and 75512, respectively. 
For conformer I, we used the transitions 5$_{1,5}-$4$_{1,4}$ and 5$_{0,4}-$4$_{0,4}$ at 31.1583057 and 32.2028617 GHz, respectively, 
For conformer II, we used the transition 6$_{2,5}-$5$_{2,4}$ at 41.9020415 GHz, respectively. These transitions are shown in Figure \ref{fig:glycineI} and \ref{fig:glycineII}, respectively.
We obtained $N<$ 8.1$\times$10$^{12}$\,cm$^{-2}$ and $N<$ 2.5$\times$10$^{11}$\,cm$^{-2}$ for glycine conformers I and II, respectively. 
The upper limits of molecular abundances are $<$ 0.6$\times$10$^{-10}$ and $<$ 0.019$\times$10$^{-10}$, respectively (Table \ref{tab:isomers}). The upper limit derived for the high-energy conformer II is much lower because its dipole moment is larger by a factor of 5 (see Table \ref{tab:isomers}). If both conformers were in thermodynamic equilibrium, from the upper limit of the conformer II we could predict the expected column density of conformer I. Using their relative energy ($\Delta E$= 403 K, see Table \ref{tab:isomers}), and assuming $T_{\rm k}=$ 100 K, the ratio between the conformers would be $\sim$56. Using this ratio, the upper limit of the low-energy conformer I would be $\sim$1$\times$10$^{-10}$, very similar to that derived directly using conformer I transitions (Table \ref{tab:isomers}). 

\begin{figure}
\begin{center}
\includegraphics[width=9cm]{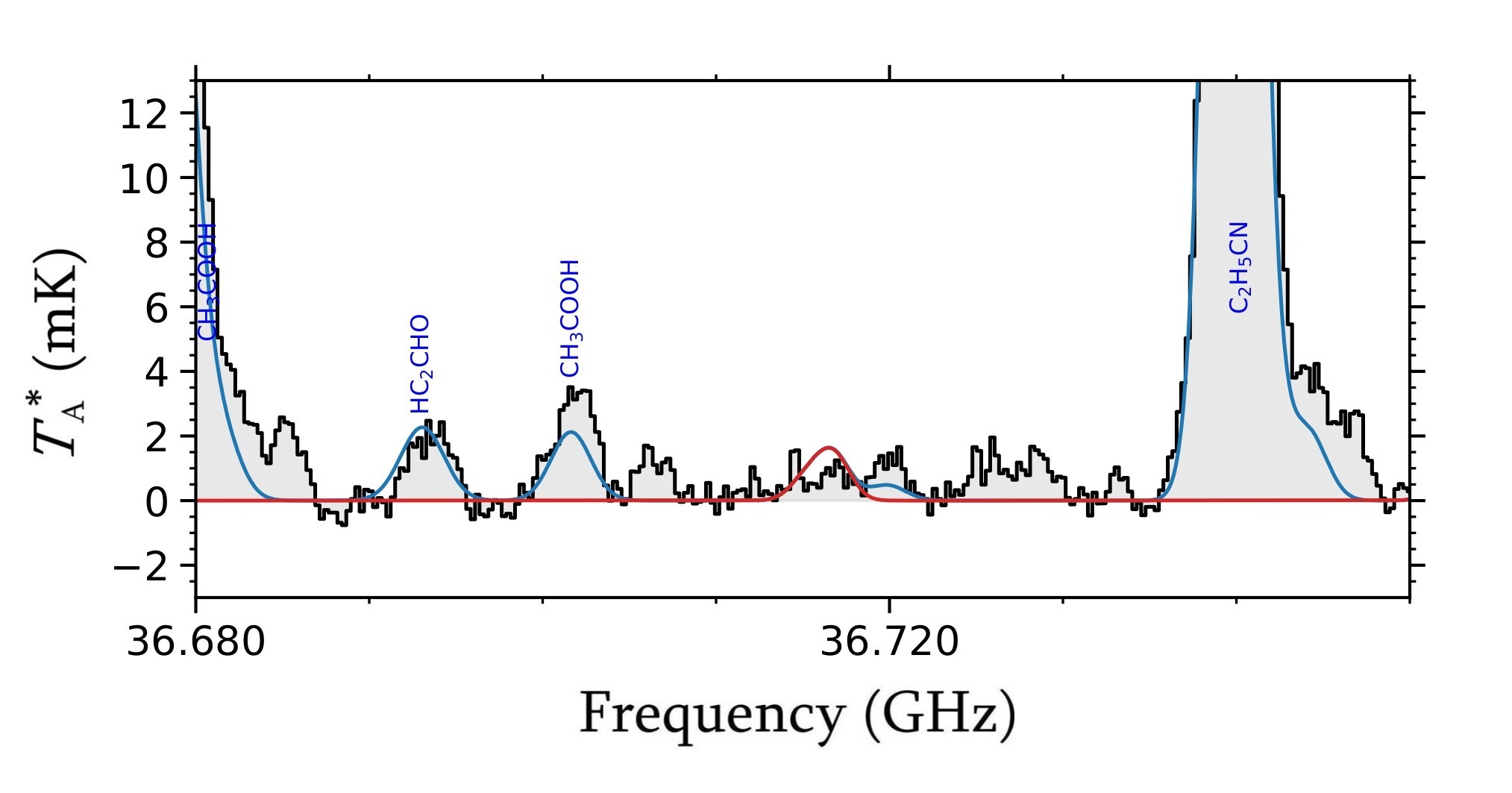}
\end{center}
\vspace{-5mm}
\caption{Transition of $syn-$methyl carbamate (\ch{CH3OC(O)NH2}) used to derive its column density upper limit. The observed spectra are shown as a grey histogram, and the LTE synthetic model using the derived upper limit is shown with a red curve.}
\label{fig:methhylcarbamate}
\end{figure}

$Syn-$Methyl carbamate (\ch{CH3OC(O)NH}) is the lowest-energy \ch{C2H5O2N} isomer for which rotational spectroscopy is available (Table \ref{tab:isomers}). To compute its abundance upper limit, we used the Jet Propulsion Laboratory catalog (JPL; \citealt{pickett1998}) entry 75004. 
We used the five hyperfine components of the 2$_{2,0}-$1$_{1,1}$ transition at $\sim$36.72 GHz (Figure \ref{fig:methhylcarbamate}), which provides $N<$ 3.3$\times$10$^{13}$\,cm$^{-2}$, i.e. a molecular abundance of $<$ 2.5$\times$10$^{-10}$ (Table \ref{tab:isomers}).

\begin{figure}
\begin{center}
\includegraphics[width=9cm]{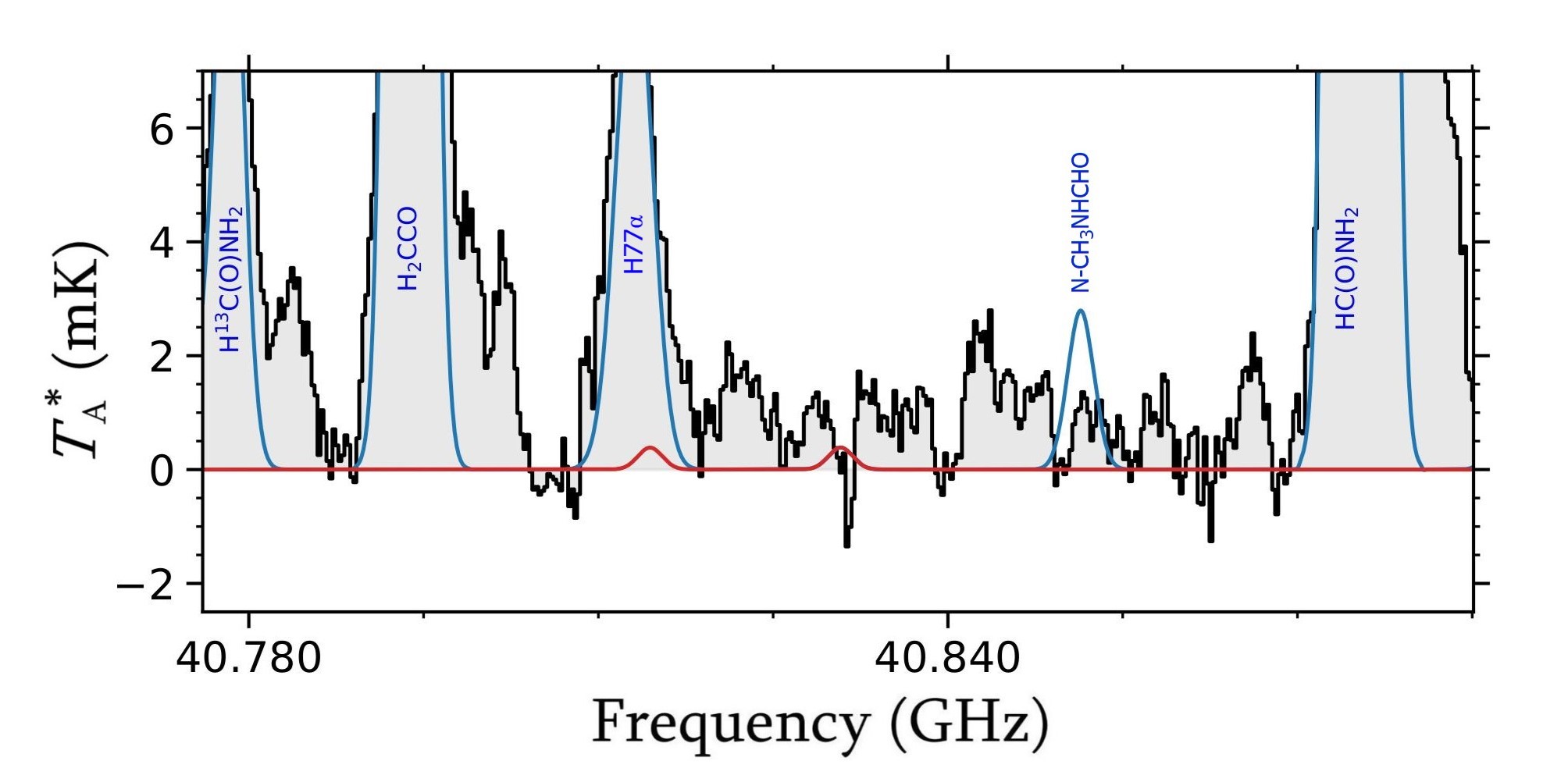}
\end{center}
\vspace{-5mm}
\caption{Transition of N-(Z)-hydroxyacetamide (\ch{CH3C(O)NHOH}) used to derive its column density upper limit. The observed spectra are shown as a grey histogram, and the LTE synthetic model using the derived upper limit is shown with a red curve.}
\label{fig:hydroxyacetamide}
\end{figure}

Similar to glycolamide, N-hydroxyacetamide (or acetohydroxamic acid, \ch{CH3CONHOH}) is a derivative of acetamide, in which an H atom of the amine group is replaced by the hydroxyl group.
Its lower-energy conformer is the Z-conformer, whose spectroscopy was presented in \citet{sanz-novo2022b}. Although an upper limit towards G+0.693 was also reported in this work, we recalculate it here using the new deeper observations, and using the same values of $T_{\rm ex}$, $v_\text{LSR}$, and FWHM values used for the other \ch{C2H5O2N} isomers. To derive the upper limit to the column density, we have used the 6$_{1,6}-$5$_{0,5}$ transition at 40.8305064 GHz (Figure \ref{fig:hydroxyacetamide}), which was also used by \citet{sanz-novo2022b}. We obtained $N<$ 2.3$\times$10$^{13}$\,cm$^{-2}$, i.e. a molecular abundance of $<$ 1.7$\times$10$^{-10}$ (Table \ref{tab:isomers}), slightly lower than that reported by \citet{sanz-novo2022b} using previous observations and assuming $T_{\rm ex}$ = 8 K, and FWHM = 20 km s$^{-1}$.

\end{document}